\documentclass[aps,prb,twocolumn,superscriptaddress]{revtex4}
\usepackage{pifont}
\usepackage{txfonts}
\usepackage{amssymb}
\usepackage{graphicx}

\begin{document}

\title{Phase diagram of a Bose-Fermi mixture in a one-dimensional optical lattice
in terms of fidelity and entanglement}
\author{Wen-Qiang Ning}
\affiliation{Department of Physics and Institute of Theoretical
Physics, The Chinese University of Hong Kong, Shatin, Hong Kong,
China} \affiliation{Department of Physics, Fudan University,
Shanghai 200433, China}
\author{Shi-Jian Gu}
 \email{sjgu@phy.cuhk.edu.hk}
\affiliation{Department of Physics and Institute of Theoretical
Physics, The Chinese University of Hong Kong, Shatin, Hong Kong,
China}
\author{Chang-Qin Wu}
\affiliation{Department of Physics, Fudan University, Shanghai
200433, China}
\author{Hai-Qing Lin}
\affiliation{Department of Physics and Institute of Theoretical
Physics, The Chinese University of Hong Kong, Shatin, Hong Kong,
China} \affiliation{Department of Physics, Fudan University,
Shanghai 200433, China}

\date{\today}

\begin{abstract}
We study the ground-state phase diagram of a Bose-Fermi mixture
loaded in a one-dimensional optical lattice by computing the
ground-state fidelity and quantum entanglement. We find that the
fidelity is able to signal quantum phase transitions between the
Luttinger liquid phase, the density-wave phase, and the phase
separation state of the system; and the concurrence can be used to
signal the transition between the density-wave phase and the Ising
phase.
\end{abstract}

\maketitle

Ultra-cold atomic gas loaded in optical lattice are attracting more
and more attentions due to ambitions of getting deep insight into
some essential physical phenomenons, such as quantum phase
transitions (QPTs), \cite{sachdev} in the condensed matter physics.
The prediction \cite{jaksch} and the successful observation
\cite{greiner} of the QPT from a superfluid to a Mott-insulator was
one of the greatest works in this field. Quite recently,
experimental progresses are very promising for studying more
non-trivial quantum phases in clod atomic systems. For example,
bosonic and fermionic atoms can simultaneously be trapped in optical
lattice in a controllable way. This ultra-cold atomic system, the
so-called Bose-Fermi mixture, often reminds people of the solid
state systems including the electron-phonon interaction. The latter
have been studied for a long period and have a complicated quantum
phase diagram. So to explore new possible phases in this system
becomes an interesting theoretical problem. Along this line, several
works had been done. \cite{das,cazalilla,buchler,lewenstein,pollet}
It is worthwhile to mention that Lode Pollet \emph{et. al.} recently
studied the ground-state phase diagram of a Bose-Fermi mixture
loaded in a one-dimensional (1D) optical lattice by using quantum
Monte Carlo (QMC) simulations. \cite{pollet} Several phases,
including Luttinger liquid (LL) phase, density wave (DW) phase,
phase separation (PS) state, and Ising phase, are predicted.

In recent years, some concepts emerging in the quantum information
theory \cite{nilesen} are extensively used to study the critical
phenomena in quantum many-body systems. One of the typical examples
is the entanglement. Many efforts have been made to the relation
between the entanglement and QPTs.
\cite{AOsterloh02,SJGu03,reviewentqpts} Quite recently, the
fidelity, as a measure of similarity between states, was proposed to
study the critical phenomena. The motivation is very simple: a
dramatic change in the structure of the ground state around the
quantum critical point should result in a great difference between
the two ground states on the both sides of the critical point. The
fidelity has been successfully applied to study the spin, fermionic,
and most recently bosonic systems.
\cite{quan,zanardi,zanardi2,buonsante,you,zhou,LCVenuti07,SChen07,MFYang07}
Compared with entanglement, the fidelity is purely a geometrical
quantity; an obvious advantage is that in analyzing the QPTs it does
not require a priori knowledge of the order parameter and the
symmetry of the system.

In this paper we try to study the ground-state fidelity and the
entanglement of the Bose-Fermi mixture in a 1D optical lattice. The
aim is two-folded: one is to test the role of fidelity and
entanglement in a more realistic system, the other is to study the
ground-state properties of the Bose-Fermi mixture. Follow Lode
Pollet \emph{et. al.}\cite{pollet}, we assume that a mixture of
bosonic and fermionic atoms is loaded into a 1D optical lattice, and
the temperature is low enough such that quantum degeneracy is
achieved. The system is then described by a lowest-band Bose-Fermi
Hubbard model,
\begin{eqnarray}
H&=&-\sum\limits_{i=1}^{N}(t_{\rm{F}}c_{i}^\dag
c_{i+1}+t_{\rm{B}}b_{i}^\dag b_{i+1}+\rm{H.c.}) \nonumber
\\&&
+U_{\rm{BF}}\sum\limits_{i=1}^{N}c_{i}^{\dag}c_{i}b_{i}^{\dag}b_{i}
+U_{\rm{BB}}\sum\limits_{i=1}^{N}b_{i}^{\dag}b_{i}(b_{i}^{\dag}b_{i}-1),
\end{eqnarray}
where $b_i$ ($b_i^+$) and $c_i$ ($c_i^+$) are the bosonic and
fermionic annihilation (creation) operators at site $i$,
respectively. Bosons (fermions) can hop from site $i$ to the nearest
neighbor site $i \pm 1$ with tunneling amplitude $t_{\rm{B}}$
($t_{\rm{F}}$). Furthermore, a large occupation of bosons on a
single site is suppressed by the on-site repulsion interaction
$U_{\rm{BB}}$. Bosons and fermions can mutually repel or attract
each other on each site depending on the sign of $U_{\rm{BF}}$. In
this paper, we choose $U_{\rm{BF}} > 0$, and consider the case where
both of the bosons and the fermions have a density as:
$N_{\rm{F}}=N_{\rm{B}}=N/2$.

We now briefly introduce the ground-state phase diagram of the
model. When the $U_{\rm{BF}}$ is small enough, the fermions behave
as a LL, the interaction between them is induced by the bosons. At
the same time, the bosons form an interacting liquid too. So we have
a LL of fermions, which weakly interacts a boson liquid. When
$U_{\rm{BB}}$ is small and $U_{\rm{BF}}$ is large, the system is in
the first order unstable to the PS with hard domain walls; in other
words the system is separated into two regions: a bosonic region and
a fermionic region. When both $U_{\rm{BB}}$ and $U_{\rm{BF}}$ are
very large, the bosons behave as fermions, which means that the
occupation of more than one boson at a single site is not allowed.
The model can then be mapped into the 1D XXZ model: $H_{XXZ} =
\sum_i J(\sigma_i^x\sigma_{i+1}^x + \sigma_i^y\sigma_{i+1}^y) +
J^z\sigma_i^z\sigma_{i+1}^z$, where
$J=-(t_{\rm{B}}t_{\rm{F}})/U_{\rm{BF}}$ and
$J^z=(t_{\rm{B}}^2+t_{\rm{F}}^2)/(2U_{\rm{BF}})-
t_{\rm{B}}^2/(2U_{\rm{BB}})$ \cite{duan}. This implies that there
are three phases in this limit: the ferromagnetic phase, a gapless
DW phase and a gapped Ising phase. In the ferromagnetic phase,
boson-boson bonds and fermion-fermion bonds are favored compared
with bonson-fermion bonds due to the larger exchange interactions,
this mechanism makes the system form two regions with hard domain
walls. So the ferromagnetic phase corresponds to the PS in the
mixture. This mechanism is similar to the one appearing in the 1D
asymmetric Hubbard model \cite{giamarchi,gu} where the PS is also
occurred when the system is away from half-filling. While in the DW
phase and Ising phase, the system always favors bonson-fermion
bonds. We could like to emphasis that the PS in the large
$U_{\rm{BB}}$ limit and the one in the small $U_{\rm{BB}}$ limit are
different since that the latter allows the occupation of more than
one boson on a single site. In the whole PS region, with the
increasing of $U_{\rm{BB}}$, the boson repulsion exerts a pressure
such that the region occupied by the bosons will grow and at the
same time the local density of bosons will decrease.

As mentioned, the fidelity is nothing but the modulus of the overlap
of two ground states relative to two different choices of the
Hamiltonian parameters. In this paper, we mainly focus on these two:
\begin{eqnarray}
F(2\delta U_{\rm{BF}},U_{\rm{BF}},U_{\rm{BB}}) = |\langle
\psi_{U_{\rm{BF}} - \delta U_{\rm{BF}}}|\psi_{U_{\rm{BF}} + \delta
U_{\rm{BF}}} \rangle|, \\
F(2\delta U_{\rm{BB}},U_{\rm{BF}},U_{\rm{BB}}) = |\langle
\psi_{U_{\rm{BB}} - \delta U_{\rm{BB}}}|\psi_{U_{\rm{BB}} + \delta
U_{\rm{BB}}} \rangle|,
\end{eqnarray}
in which $|\psi_\lambda\rangle$ stands for the ground state of the
Hamiltonian (1) with the parameter $\lambda$, and is calculated by
the Lanczos method for a finite sample. To avoid the ground-state
level crossing, anti-periodic boundary conditions (APBCs) are
applied for system size $N = 4n$ and periodic boundary conditions
(PBCs) for $N = 4n+2$, where $n$ is an integer. According to the
original motivation of the fidelity, a drop in the fidelity of two
ground states separated by two lightly different parameters is
expected to be a signature of the QPT.

\begin{figure}
\includegraphics[scale=0.35]{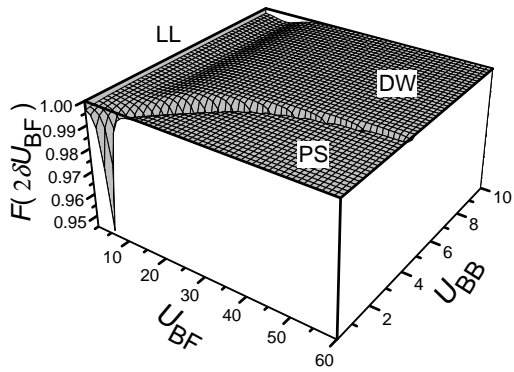}
\caption{The obtained fidelity $F(2\delta
U_{\rm{BF}},U_{\rm{BF}},U_{\rm{BB}})$ when $t_{\rm{F}} = 4.0,
t_{\rm{B}} = 1.0, \delta U_{\rm{BF}} = 0.5$. It can be observed
obviously that there are two phase transition boundary lines
indicated by the drop in the fidelity. The phases such as LL, DW,
and PS are identified by the comparing this phase diagram with the
one which was proposed earlier by the calculation of the correlation
functions (see text). $N = 8$ (APBC).}
\end{figure}

In Fig. 1, we show one of our main results, i.e. the fidelity
$F(2\delta U_{\rm{BF}},U_{\rm{BF}},U_{\rm{BB}})$ defined on the
$U_{\rm {BB}}-U_{\rm{BF}}$ plane. Compare this figure with Fig. 4 of
Ref. 8, perfect similarity can be observed. The boundary lines of
phase transitions between the LL, the DW and the PS are clearly
indicated by the drop of the fidelity. We would like to emphasis
that the phase diagram presented here is obtained in such a small
cluster and without any knowledge of the correlation properties of
the system. It is also clearly observed that the drop of the
fidelity along the phase transition line between the DW and PS
phases becomes deeper and deeper as the interaction decreases. This
phenomenon indicates that although the phase transition is within
the same class but the similarity of the ground state is changing
along this line.

However, the phase transition, as reported in Ref. 8, between the DW
and the Ising phases is not indicated in Fig. 1. According to the
effective model, i.e. 1D XXZ model, this transition belongs to the
KT universality class. \cite{pollet,fach} Lode Pollet \emph{et. al.}
\cite{pollet} did numerical calculation of the correlation functions
in a relatively larger system ($N \sim 30$) by the QMC simulations
and claimed that a true long-range order may exist in the Ising
phase. It is highly difficult for us to make a scaling analysis of
the correlation functions by the Lanczos method. However, the 1D XXZ
\cite{yang} model provides us a clue to investigate this problem. It
was reported that the concurrence, \cite{wootters} as a measure of
entanglement between two qubits, reaches maximum at the SU(2) point
\cite{SJGu03} of the XXZ model. This maximum point corresponds to
the transition point between the DW and the Ising phases.

The concurrence in the spin models can be calculated in the
following way. Due to the global SU(2) symmetry of the XXZ model,
the $z$-component of the total spin of the system is a good quantum
number, the reduced density matrix $\rho_{i,i+1}$ of two neighboring
spins has the form
\begin{eqnarray}
\rho _{i,i + 1}  = \left( {\begin{array}{*{20}c}
   {u^ +  } & 0 & 0 & 0  \\
   0 & w & {z^* } & 0  \\
   0 & z & w & 0  \\
   0 & 0 & 0 & {u^ -  }  \\
\end{array}} \right),
\end{eqnarray}
in the spin basis {$|\uparrow \uparrow \rangle$, $|\downarrow
\uparrow \rangle$, $|\uparrow \downarrow \rangle$, $|\downarrow
\downarrow \rangle$}. The elements in the reduced density matrix
$\rho_{i,i+1}$ can be obtained from the correlation functions
\begin{eqnarray}
u^{\pm} =
\frac{1}{4}(1\pm2\langle\sigma_i^z\rangle+\langle\sigma_i^z\sigma_{i+1}^z\rangle),
\\
w = \frac{1}{4}(1-\langle\sigma_i^z\sigma_{i+1}^z\rangle),\\
z =
\frac{1}{4}(\langle\sigma_i^x\sigma_{i+1}^x\rangle+\langle\sigma_i^y\sigma_{i+1}^y\rangle
+i\langle\sigma_i^x\sigma_{i+1}^y\rangle-i\langle\sigma_i^y\sigma_{i+1}^x\rangle).
\end{eqnarray}
All the information needed is contained in the reduced density
matrix, from which the concurrence is readily obtained as
\cite{wang}
\begin{eqnarray}
C = 2\max \left[ {0,\left| z \right| - \sqrt {u^ +  u^ -  } }
\right],
\end{eqnarray}
which can be expressed in terms of the correlation functions and the
magnetization by using equation (5), (6) and (7).

The concurrence is only valid for two-qubit system. For the
Bose-Fermi Hubbard model, the double occupation of two particles is
almost not allowed, the state of the two neighboring sites can be
described by four basis {$|ff\rangle$, $|bf\rangle$, $|fb\rangle$,
$|bb\rangle$} where $f$ ($b$) represents there is only one fermion
(boson) on a single site. So we can associate a pseudo-spin ``up''
(``down'') with the ``$f$'' (``$b$''). Then the equations described
above can also be perfectly applied to the Bose-Fermi Hubbard model.
In other words, if we calculate the trace of the reduced matrix of
two nearest sites, which reads $T = \rm{Tr}\rho_{i,i+1}$, then $T$
must equal to $1$ in this limit. While $T$ only approximately equals
$1$, when $U_{\rm{BB}}$ and $U_{\rm{BF}}$ are not infinite but
large. As long as $(1-T)$ is small enough the concurrence calculated
by Eq. (8) is a good characterization of entanglement between two
sites. In this way, we are able to calculate the concurrence in the
Bose-Fermi mixtures approximately. This kind of treatment was also
used by some other group recently\cite{kgh}. Obviously, as expected
a maximum is clearly observed in the behavior of the concurrent,
which indicates a QPT point. It is important to point out that the
variety of the concurrence is not a dramatic one in the whole DW
region. Using the exact solution of the XXZ chain, the phase
transition from the DW phase to the Ising phase should occur at
$-J=J^z$, i. e. $U_{\rm{BB}}=6.7$ for $t_{\rm{F}} = 4.0, t_{\rm{B}}
= 1.0, U_{\rm{BF}}=60$. But according to the obtained concurrence,
the max point $U_{\rm{BB}}^{max}$ approximately equals to $7.7$. The
discrepancy comes from the approximations made and the size effect.

\begin{figure}
\includegraphics[scale=0.35]{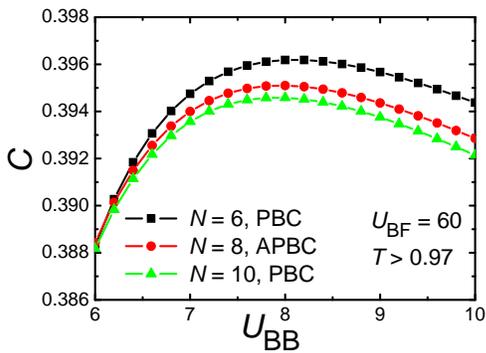}
\caption{The obtained concurrence in the DW phase where $t_{\rm{F}}
= 4.0, t_{\rm{B}} = 1.0$. The concurrence of the $N=6$ ($N=8$) had
been reduced by 0.029 (0.009) in order to make the results reside in
the same region in figure. All other parameters are indicated in the
figure.}
\end{figure}

Some skeptical readers may wonder why other kinds of measurements of
the entanglement, for example the von Neumann entropy, are not
calculated here since these kinds of measurements are more suitable
for the Bose-Fermi Hubbard model at the first glance of their
definitions. We actually calculated von Neumann entrpy though we did
not show here. But it turns out that these quantities are more
difficult to witness the phase transition between the DW and Ising
phase.

After all the analysis, we contribute the missing of the phase
transition signature between the DW phase and the Ising phase in the
fidelity to two reasons. The first one is that this kind of
transition is actually a very weak one, which means that the change
of the ground state around the critical point is not dramatic, at
least in the finite size system according to our results. The second
one is that, as reported before, the fidelity may not be a good
indicator of those transitions of infinite order, such as KT
transitions. \cite{you} However, further studies are definitely
needed in order to answer the question completely!

\begin{figure}
\includegraphics[scale=0.35]{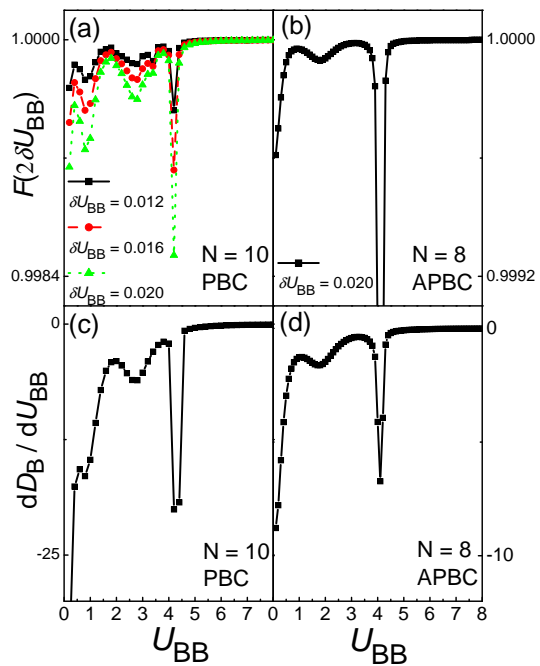}
\caption{The obtained fidelity $F(2\delta
U_{\rm{BB}},U_{\rm{BF}},U_{\rm{BB}})$ (a), (b) and the first order
derivative of the local density of boson (see text)(c), (d) when
$t_{\rm{F}} = 4.0, t_{\rm{B}} = 1.0, U_{\rm{BF}} = 60.0$. All other
parameters are indicated in the figure.}
\end{figure}

We also calculated the fidelity $F(2\delta
U_{\rm{BB}},U_{\rm{BF}},U_{\rm{BB}})$ which can signal the
transition between the PS phase and DW phase too. As you can see in
the Fig. 3, the most dramatic drop in the fidelity is in
correspondence with the phase transition point between the PS state
and the DW phase. The critical point find here is consistence with
the one found in the Fig. 1. In addition, some other drops in the
fidelity can be observed in the PS region. These drops are related
to the changing rate of the local density of boson. To show this, we
define the local density of boson as,
\begin{eqnarray}
D_{\rm{B}}
=\langle\frac{1}{N}\sum\limits_{i=1}^{N}b_{i}^{\dag}b_{i}(b_{i}^{\dag}b_{i}-1)\rangle,
\end{eqnarray}
and its first order derivative is calculated. Comparing Fig. 3 (a)
and Fig. 3 (b) with Fig. 3 (c) and Fig. 3 (d) respectively, every
drop in the fidelity has its company, a drop in the derivative of
the local density of boson. This relation can be understood by the
dramatic change of the local density of boson will lead to a big
change of the ground state wave-function, and then will make the
fidelity decrease greatly. The drop of the fidelity at the QPT point
between the PS state and DW phase can also be thought like this way.
These observations strongly imply that the transition between PS
phase and DW phase is within the Landau's symmetry breaking theory
and a first order one. Furthermore, it is easy to notice that the
phase transition point is not affected by the system size and the
boundary conditions used, but the drops found in the PS phase is
strongly affected, which indicates that these drops may be a size
effect and can not be identified as phase transitions. Furthermore,
the transition between DW phase and Ising phase is not observed
again!

In summary, we calculated the fidelity $F(2\delta
U_{\rm{BF}},U_{\rm{BF}},U_{\rm{BB}})$ and $F(2\delta
U_{\rm{BB}},U_{\rm{BF}},U_{\rm{BB}})$ of the 1D Bose-Fermi Hubbard
model, which can be used to describe low-temperature physics of the
atomic Bose-Fermi mixtures loaded in 1D optical lattices. It is
showed that although the ground state phase diagram of this system
is complicated, and the fidelity may be a good tool to study the
phase transitions without a priori knowledge of the order parameter
and the symmetry of the system. But one should be very careful
because the fidelity may fail for the transition which is belong to
the KT-like type, for example, the transition between DW phase and
Ising phase in this system. We also calculate the concurrence in the
DW phase. The result indicate a QPT may exist in the DW phase
although the fidelity have no singulary at this transition point.

\begin{acknowledgements}

This work was partially supported by the National Natural Science
Foundation of China, RGG Grant CUHK 400906, 401504, and MOE B06011.
W.Q.N. would like to thank the support of the Graduated Students'
Innovation Foundation of Fudan University.

\end{acknowledgements}

\end{document}